\documentclass[final,5p,times,twocolumn]{elsarticle}  

\biboptions{compress}
\usepackage[T1]{fontenc}

\usepackage{amssymb, amsmath , lipsum, xspace, graphicx, listings, epsfig, caption, subcaption}
\usepackage{natbib}
\usepackage[dvipsnames, usenames]{xcolor}

\definecolor{codegreen}{rgb}{0,0.6,0}
\definecolor{codegray}{rgb}{0.5,0.5,0.5}
\definecolor{codepurple}{rgb}{0.58,0,0.82}
\definecolor{backcolour}{rgb}{1.,1.,1.}
\lstdefinestyle{codestyle}{
    language=Python,
    backgroundcolor=\color{backcolour},   
    commentstyle=\color{codegreen},
    keywordstyle=\color{magenta},
    numberstyle=\tiny\color{codegray},
    stringstyle=\color{codepurple},
    basicstyle=\ttfamily\footnotesize,
    breakatwhitespace=false,         
    breaklines=true,                 
    captionpos=b,                    
    keepspaces=true,                 
    numbers=none,                    
    numbersep=5pt,                  
    showspaces=false,                
    showstringspaces=false,
    showtabs=false,                  
    tabsize=2
}

\definecolor{linkcolor}{rgb}{0.0,0.3,0.5}
\usepackage[unicode, colorlinks=true, linkcolor=linkcolor, citecolor=linkcolor, filecolor=linkcolor, urlcolor=linkcolor]{hyperref}
\usepackage{orcidlink}

\newcommand{\elhs}{\textsc{expandLHS}\xspace}
\newcommand{\python}{\textsc{Python}\xspace}
\newcommand{\git}{\textsc{git}\xspace}
\newcommand{\numpy}{\textsc{Numpy}\xspace}
\newcommand{\scipy}{\textsc{Scipy}\xspace}
\newcommand{\numba}{\textsc{Numba}\xspace}
\newcommand{\D}{\mathcal{D}\xspace}

\journal{Software X}

\begin{document}

\begin{frontmatter}



\title{``LHS in LHS'': a new expansion strategy for Latin hypercube sampling\\in simulation design}


\author[a,b]{Matteo Boschini \orcidlink{0009-0002-5682-1871}}
\ead{m.boschini1@campus.unimib.it}
\author[a,b]{Davide Gerosa \orcidlink{0000-0002-0933-3579}}
\author[c]{Alessandro Crespi \orcidlink{0009-0008-2475-1569}}
\author[a]{Matteo Falcone \orcidlink{0009-0009-5961-6698}}
\affiliation[a]{organization={Dipartimento di Fisica ``G. Occhialini'', Universit\'a degli Studi di Milano-Bicocca},
            addressline={Piazza della Scienza 3}, 
            city={Milano},
            postcode={20126}, 
            country={Italy}}
\affiliation[b]{organization={INFN, sezione di Milano-Bicocca},
            addressline={Piazza della Scienza 3}, 
            city={Milano},
            postcode={20126}, 
            country={Italy}}
\affiliation[c]{organization={Dipartimento di Informatica, Sistemistica e Comunicazione, Universit\'a degli Studi di Milano-Bicocca},
            addressline={Viale Sarca 336}, 
            city={Milano},
            postcode={20126}, 
            country={Italy}}

\begin{abstract}
Latin Hypercube Sampling (LHS) is a prominent tool in simulation design, with a variety of applications in high-dimensional and computationally expensive problems. LHS allows for various optimization strategies, most notably to ensure space-filling properties. However, LHS is a single-stage algorithm that requires a priori knowledge of the targeted sample size.
In this work, we present ``LHS in LHS’’, a new expansion algorithm for LHS that enables the addition of new samples to an existing LHS-distributed set while (approximately) preserving its properties. In summary, the algorithm identifies regions of the parameter space that are far from the initial set, draws a new LHS within those regions, and then merges it with the original samples.
As a by-product, we introduce a new metric, the LHS degree, which quantifies the deviation of a given design from an LHS distribution. Our public implementation is distributed via the Python package \elhs.


\bigskip \hrule \bigskip
\newpage
\noindent{\bf Code metadata} \medskip

\noindent \hspace{-0.19cm}\begin{tabular}{p{0.52\textwidth}p{0.45\textwidth}}
Current code version & v1.1 \\
Permanent link to code/repository used for this code version & \url{https://github.com/m-boschini/expandLHS} \\
Legal Code License   &  MIT  \\
Code versioning system used & git \\
Software code languages, tools, and services used &  python \\
Compilation requirements, operating environments \& dependencies & numpy, scipy, numba (>0.57) \\
If available Link to developer documentation/manual & \url{https://m-boschini.github.io/expandLHS} \\
Support email for questions & m.boschini1@campus.unimib.it %
\end{tabular}

\end{abstract}

\end{frontmatter}

\section{Motivation and significance}

The continuous growth of available data is a common trend across disciplines, often enabling the study of increasingly complex phenomena. A useful strategy to address these challenges is to perform simulated experiments, where one analyzes the outcomes of a model and/or the predictions of a theory while controlling the effects of input parameters. Simulations are widely used in various fields, ranging from mathematics to physics, medicine, economics, and education \cite{Mittal2017}. More often than not, simulating complex phenomena carries a high computational cost, making it critical to determine where to place simulations in the parameter space.

To this end, simulation design is a branch of statistics that aims to optimize computational experiments \cite{774062c9f24141d19c01fb033aa107e5, GARUD201771}. Simulation design assists in developing algorithmic strategies that explore the parameter space effectively, ensuring an optimal distribution of samples (where optimality needs to be appropriately defined), while mitigating computational costs. A set of samples, where simulations will be placed, can therefore be preferred over another by enforcing specific properties such as space-filling, one-dimensional projection, and low correlation.
There is a vast literature on simulation design, with a variety of implementations and model-independent algorithms based on different notions of optimality \cite{774062c9f24141d19c01fb033aa107e5}. In this paper, we focus on one such approach: Latin hypercube sampling (LHS)\cite{2022arXiv220306334D}. %

LHS was first introduced in the 1970s~\cite{eglajs1977new, ef76b040-2f28-37ba-b0c4-02ed99573416} and was further developed to optimize projection properties, improve space-filling design, and avoid spurious correlations~\cite{Iman01011980, FLORIAN1992123, Tang01121993, PARK199495, MORRIS1995381, Ye01121998, YE2000145, Cioppa01022007}. The concept of Latin square comes from combinatorial mathematics --- an $N \times N$ square with $N$ different symbols appearing only once per column and row. A Latin hypercube generalizes this property to a $P$-dimensional hypercube, where each dimension is binned into $N$ disjoint intervals $[i/N, (i+1)/N)$ where $i = 0,1,...,N-1$ with marginal probability $1/N$. LHS distributes points to tile the space while preserving the one-dimensional projection property—i.e., ensuring that each marginalized one-dimensional bin contains exactly one sample. Multiple designs with this property are possible for a given $N$, and one can further optimize the final result by imposing additional criteria, such as orthogonality and distance optimization \cite{2022arXiv220306334D}. However, these are typically \textit{single-stage} sampling algorithms, requiring all samples to be drawn simultaneously or, at most, in a finite number of discrete steps. %
This implies that the number of targeted samples must be known \textit{a priori}, and once the set is drawn, additional samples generally cannot be added while preserving the desired projection property.

This is a crucial limitation in many practical applications. In most realistic scenarios, simulation design does not happen all at once but rather gradually and iteratively, with subsequent batches that need to be properly initialized.
Suppose one initially plans for 
$N$ simulations based on the available computational budget and distributes them using LHS. Later, realizing that this is insufficient for the targeted application, one might secure additional computing time for 
$M$ more simulations. How should these 
$M$ additional simulations be integrated into the initial LHS-distributed set of 
$N$ simulations so that the combined set of 
$N+M$ samples still retains desirable space-filling properties?

In this paper, we propose a new model-free expansion algorithm for LHS, accompanied by an open-source package for the \python programming language, which we dubbed  \elhs.
Compared to other proposed LHS expansion strategies~\cite{Xiong01082009, CROMBECQ2011683, vovrechovsky2015hierarchical, SHEIKHOLESLAMI2017109, Li02092017, Zhou2019}, our approach allocates new samples by rebinning the original hypercube to preserve the projection property. Building on this idea, we can further leverage the versatility of Latin hypercubes by essentially designing an LHS within the existing LHS. As a by-product, this paper introduces the notion of “LHS degree” to quantify how closely a given set of samples resembles an LHS with the same number of points.

\section{Software description}
\label{sec:two}

\subsection{Algorithm}

Let us denote by 
${\rm LHS}(P,N)$ an initial set of 
$N$ points in 
$P$ dimensions, defined within the hypercube 
$[0,1)^P$
 and satisfying the one-dimensional projection property of LHS. That is, there is one and only one sample per interval if each dimension is uniformly binned into $N$ disjoint intervals. We wish to distribute $M$ new samples to obtain an extended set ${\rm eLHS}(P, N+M)$.
 
\begin{figure*}[tb]
   \centering
   \includegraphics[width=\textwidth]{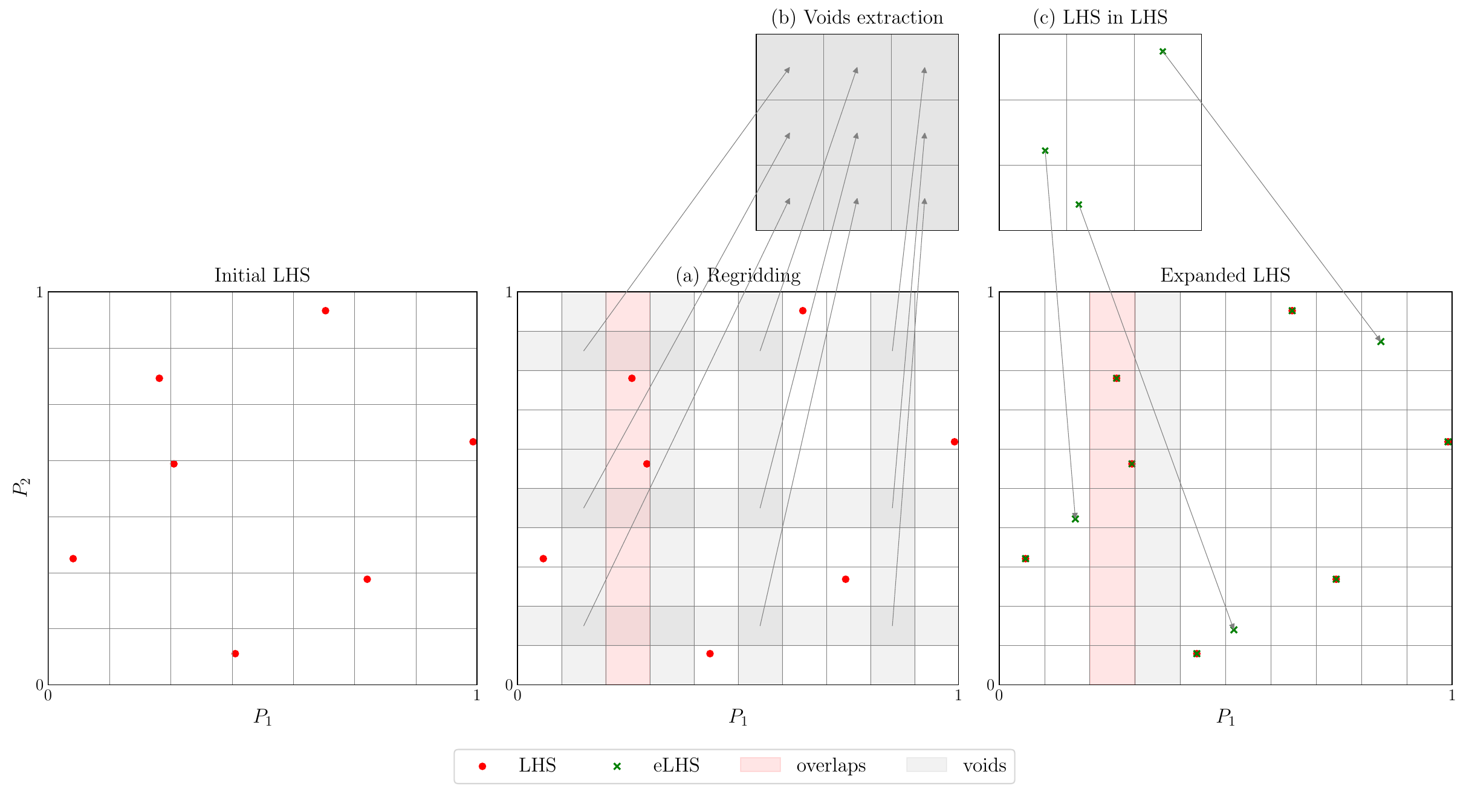} 
   \caption{The ``LHS in LHS'' expansion algorithm. The left panel shows an initial LHS with $N=7$ points in $P=2$ dimensions.  
This LHS is here expanded with $M=3$ new points.  
In step (a), we regrid the original space with $M+N$ bins in each dimension.  
This creates at least $M$ empty intervals in both dimensions (shaded in gray).  
Along dimension $P_1$, two samples overlap in the same bin (shaded in red),  
and consequently, there is a fourth empty interval.  
In step (b), $M=3$ empty intervals are randomly selected along both dimensions.  
In step (c), a new set is drawn in this subspace via LHS.  
Finally, in the right panel, the new samples are added to the original ones  
to obtain the targeted expansion ${\rm eLHS}(P, M+N)$.  
The expansion shown in this figure has a degree $\D= 0.95 < 1$ and is thus not a perfect Latin hypercube.
}
   \label{fig:eLHS}
\end{figure*}

The algorithm we propose consists of three main steps, as illustrated in Fig. \ref{fig:eLHS}: %
\begin{enumerate}
	\item[(a)] \textit{Regridding}. Each dimension is binned into $N+M$ equal-width, disjoint intervals. 
The initial samples ${\rm LHS}(P,N)$ now fall onto a different grid, 
which creates $Q \geq M$ empty bins and may lead to overlaps of samples in other bins.
	\item[(b)] \textit{Void extraction}. We consider all empty intervals and select 
$M$ of them.
	\item[(c)] \textit{``LHS in LHS''}: In this $M$-dimensional subgrid, we distribute $M$ samples using, once more, an LHS strategy. These are added to the initial LHS to obtain the targeted extended set ${\rm eLHS}(P, N+M)$.
\end{enumerate}

The probability of obtaining multiple samples in the same interval after regridding depends on both the initial set ${\rm LHS}(P,N)$ and the value of $M$. When this happens, the final extended set ${\rm eLHS}(P, N+M)$ partially loses the one-projection property. In fact, if $Q > M$ in a certain dimension, there will necessarily be $Q - M$ bins with multiple points and, consequently, empty bins after the expansion.
 
 To quantify this loss, we introduce a new metric $\D$, which we refer to as the ``LHS degree'' of a sample set. Consider a generic sample set $S$ with $N$ elements in $P$ dimensions; this is composed of real numbers $S_{ij}$ with  $i=0,\dots,N-1$ and $j=0,\dots, P-1$. We define
\begin{equation}
\label{degree}
    \D(S) = \frac{1}{N \, P }\, \sum_{j=0}^{P-1}\sum_{l=0}^{N-1}\min\left[\sum_{i=0}^{N-1}I_{\left[\frac{l}{N}, \frac{l+1}{N}\right)}(S_{ij}), 1\right]\, ,
\end{equation}
where $I$ is the indicator function defined as
\begin{equation*}
   I_{\left[a,b\right)}(x) = 
   \begin{cases}
       1  &x \in \left[a,b\right) \\
       0  &x \notin  \left[a,b\right)
   \end{cases} \,\, .
\end{equation*}
In particular, one has $0<\D(S)\leq1$ and 
\begin{equation}
\D\left(S\right)=1 \iff S = {\rm LHS}(P,N) \,,
\end{equation} 
The degree $\D$ can thus be interpreted as the fractional closeness to an LHS set. 
Essentially, this procedure checks all the $i$-intervals for each dimension $j$. If they are populated by a sample $S_{ij}$, a weight $1/(NP)$ is assigned; otherwise, the weight is 0. The metric is the sum of all the weights.

After the expansion, one has $\D({\rm eLHS}) \leq 1$. Indeed, the sums in Eq.~(\ref{degree}) trace the presence of at least one sample per interval. An interval containing overlapping samples still contributes a weight of 1 to the weighted sum described above, due to the use of the $\min$ function. For each such interval, there exists a corresponding empty interval in the same dimension that receives a weight of zero. Consequently, the degree $\D({\rm eLHS})$ is reduced.
Figure \ref{fig:eLHS} shows an example with $N=7$, $M=3$, and $P=2$. In this case, the final degree is $\D=0.95$. 
A ``perfect'' expansion with $\D({\rm eLHS})=1$ corresponds to the case where the new $N+M$ set of points is also LHS-distributed. In general, this is not always possible.

The selection of $M$ empty intervals is not unique whenever $Q < M$.  
Among the possible choices, we select the one that maximizes  
either the centered discrepancy or the geometric discrepancy \cite{774062c9f24141d19c01fb033aa107e5, ZHOU2013283}.
The former measures the uniformity of the sample set as a proxy for its space-filling properties,  
while the latter tracks the minimum Euclidean distance between samples.
In practice, we generate multiple expansions and select the one that  
minimizes (maximizes) the discrepancy (geometric discrepancy),  
up to a certain tolerance level.  
This becomes more relevant in a larger number of dimensions $P$  
and as the number of new samples $M$ approaches the initial sample size~$N$.

\subsection{Implementation}
\label{implementation}

\begin{figure*}[h]
   \centering
   \includegraphics[width=0.9\textwidth]{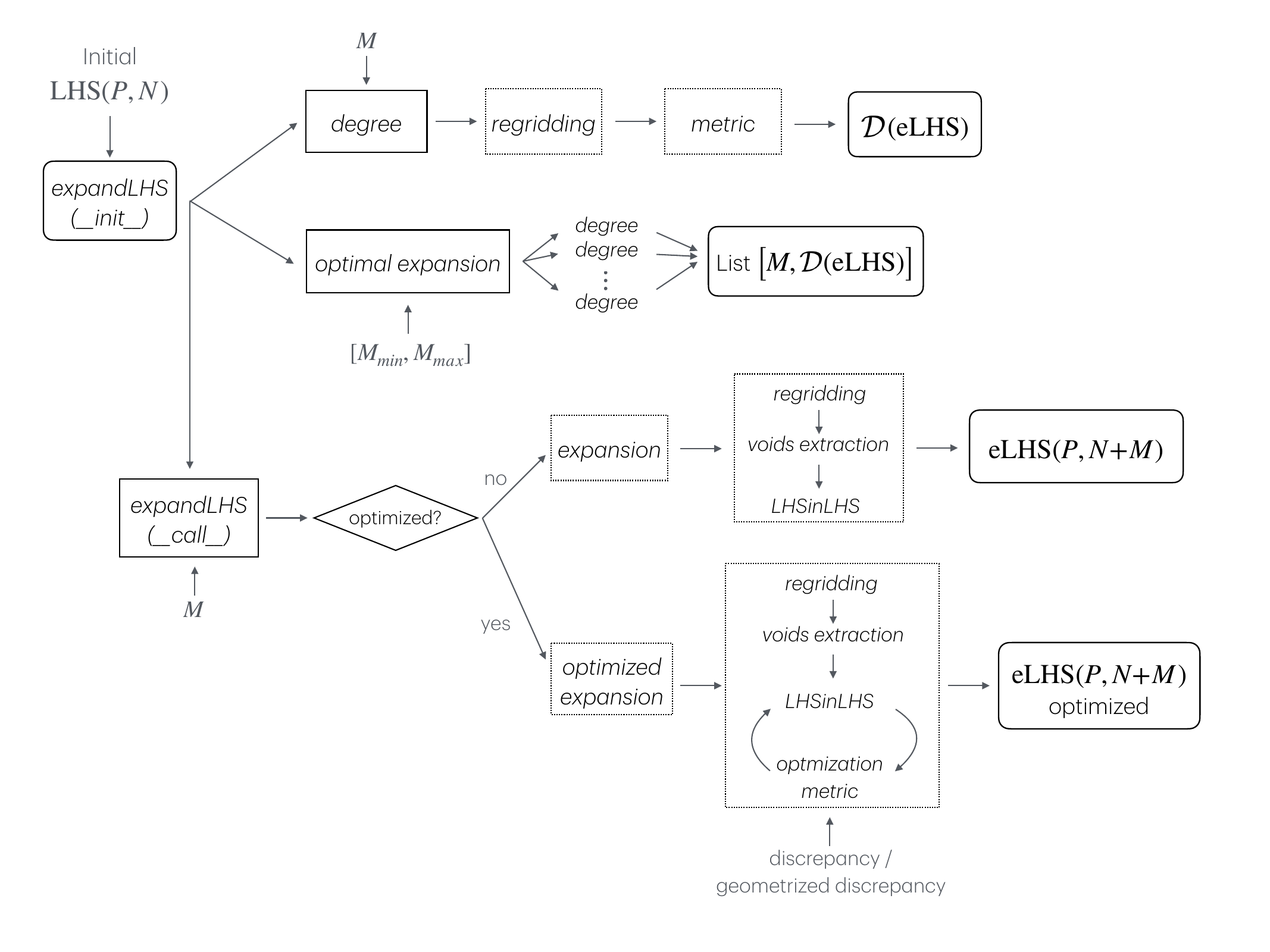} 
   \caption{Schematic representation of the \elhs package. Functionalities are implemented in the \texttt{ExpandLHS} class. Users have access to three main methods (solid black rectangles): \texttt{degree}, \texttt{optimal\_expansion}, and the \texttt{ExpandLHS()} call. These compute the degree $\D$ of a putative expansion, estimate the optimal expansion size, and perform the actual expansion, respectively. For additional details see Sec.~\ref{implementation} and the code documentation (\href{https://m-boschini.github.io/expandLHS)}{m-boschini.github.io/expandLHS). }
   }
   \label{fig:flowchart}
\end{figure*}

Our software, \elhs, is implemented in \python 3 and leverages \numpy's arrays  
and \numba's just-in-time compilation for fast computations.  
\elhs is distributed via the \python Package Index and can be installed via:
\begin{verbatim}
	pip install expandLHS 
\end{verbatim}
Dependencies include \numpy \cite{harris2020array},  \scipy \cite{2020SciPy-NMeth}, and \numba (version $\geq 0.57.0$) \cite{10.1145/2833157.2833162}. If not present, these libraries will be installed/updated along with the package.

The \elhs module is implemented as a single \python class. This is imported within a \python console, script, or Jupyter Notebook using e.g.
\begin{verbatim}
	from expandLHS import ExpandLHS
\end{verbatim}
Our software is in its v1.1 release. The source code is distributed under  
the \git version control system and the MIT permissive license at  
\href{https://github.com/m-boschini/expandLHS}{github.com/m-boschini/expandLHS}. See Fig.~\ref{fig:flowchart} for a flowchart describing the overall implementation.

The \elhs class has additional functionalities implemented. The method \texttt{ExpandLHS.degree} computes the degree $\D$ defined in Eq.~(\ref{degree}).  The method \texttt{ExpandLHS.optimal\_expansion} iterates over possible values of $M$ to identify the expansion strategy that maximizes the degree of the resulting eLHS.
Finally, the inner LHS in step (c) of the algorithm above can be  optimized to achieve a user-defined threshold in the chosen metric. At present, our package implements two different metrics from the \texttt{scipy.stats.qmc} submodule:  \texttt{discrepancy} and  \texttt{geometric\_discrepancy}.

\section{Illustrative examples}
\label{sec:three}

\subsection{LHS expansion}
The code  snippet below introduces the basic usage  of \elhs. We extend an initial LHS set with $N=20$ points in $P=2$ dimensions, adding $M=18$ new samples.

\begin{figure*}[ht]
	\centering
	{\includegraphics[width=\textwidth]{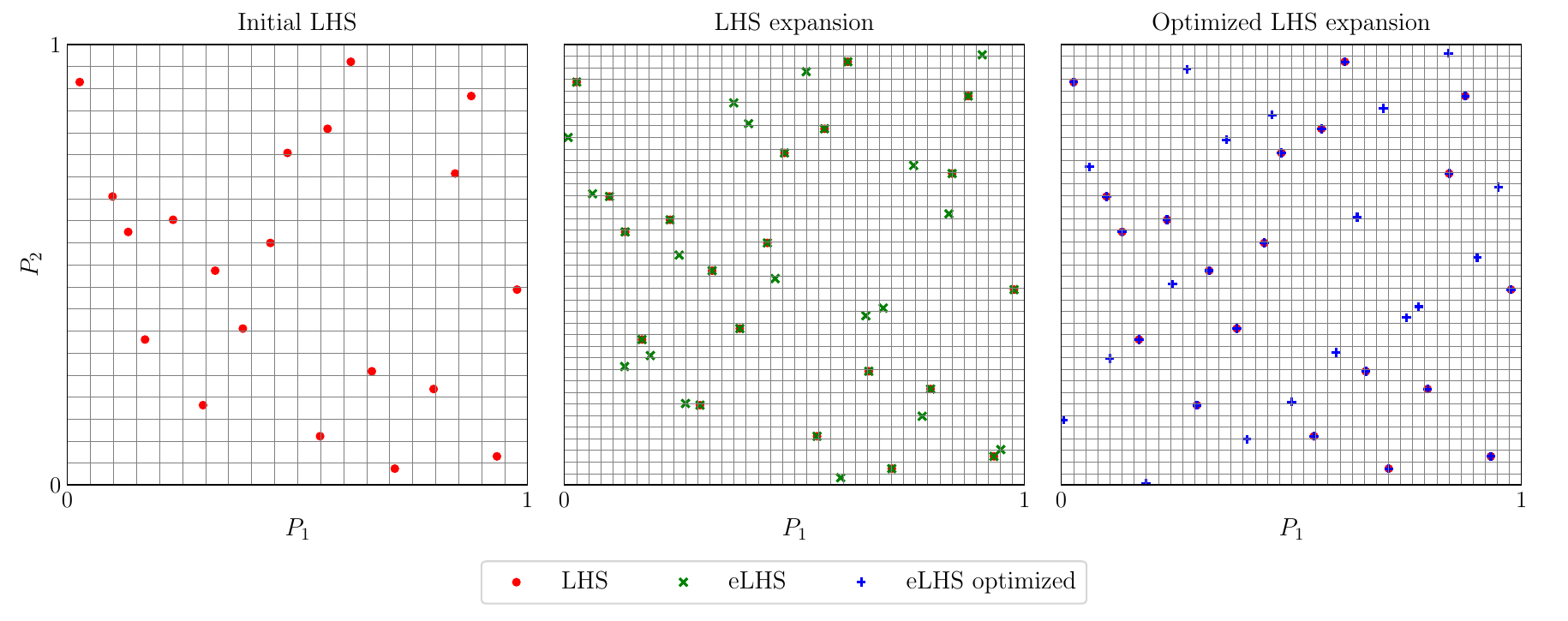}}
	\caption{Examples of LHS expansion. The left panel shows a standard LHS set with $N=20$ samples in $P=2$ dimensions (red circles). The middle panel shows a possible expansion with $M=18$ new data points (green crosses), without further optimization. The right panel shows a different expansion, which was optimized to minimize the overall discrepancy. We report centered discrepancies of $1.0\times10^{-3}$ and $4.7\times10^{-4}$ for the unoptimized and the optimized expansions, respectively.}
	\label{fig:lhs2elhs}
\end{figure*}

\begin{lstlisting}[style=codestyle]
from scipy.stats.qmc import LatinHypercube
from expandLHS import ExpandLHS
	 
# initial LHS set
P = 2 # hypercube dimension
N = 20 # initial sample size
lhs_sampler =  LatinHypercube(P)
lhs_set = lhs_sampler.random(N)
	 
# LHS expansion 
M = 18 # expansion size
eLHS = ExpandLHS(lhs_set)
elhs_set = eLHS(M)
elhs_opt = eLHS(M, optimize='discrepancy')
\end{lstlisting}
Outputs are illustrated in Fig. \ref{fig:lhs2elhs}, where we show  
the initial set, an unoptimized expanded set, and the optimized expanded set.  
Optimizing using the discrepancy criterion increases the uniformity  
of the expanded sample set. For this case, we report a discrepancy estimate  
of $1.0 \times 10^{-3}$ for the unoptimized expansion and  
$4.7 \times 10^{-4}$ for the optimized expansion. This particular case happens to be a perfect expansion, $\D({\rm eLHS}) = 1$.  

\subsection{How many new samples?}

In the following example, we simulate different expansions for the same initial set by adding a number of new points $M$ between 4 and 12, ranking our attempts according to the resulting degree $\D$. Note that this metric depends only on the initial set and the value of $M$, and not on the added samples.
\begin{lstlisting}[style=codestyle]
# optimal expansion size with M in [4, 12]
# verbose = False returns the optimal expansion
# verbose = True returns all the estimates
eLHS.optimal_expansion((4,12), verbose=True)
    M  degree 
   (0,  1.0), # no expansion
   (12, 1.0), # best option
   (9,  0.9828),
   (7,  0.9815),
   (10, 0.9667),
   (6,  0.9615),
   (11, 0.9516),
   (5,  0.92),
   (4,  0.9167),
   (8,  0.9107) # worst option
\end{lstlisting}
We obtain a perfect expansion for $M=12$, while the worst case is that with $M=8$. %
This functionality offers some interesting use cases. In the example above,  
suppose one has budgeted computational time for $M=8$ new simulations.  It turns out a minor increase to $M=9$ ensures substantially better coverage  
of the parameter space, with $\D$ increasing from $\sim 0.91$ to $\sim 0.98$.

\subsection{General behavior of eLHS}
\begin{figure*}[ht]
	\centering
	{\includegraphics[width=\textwidth]{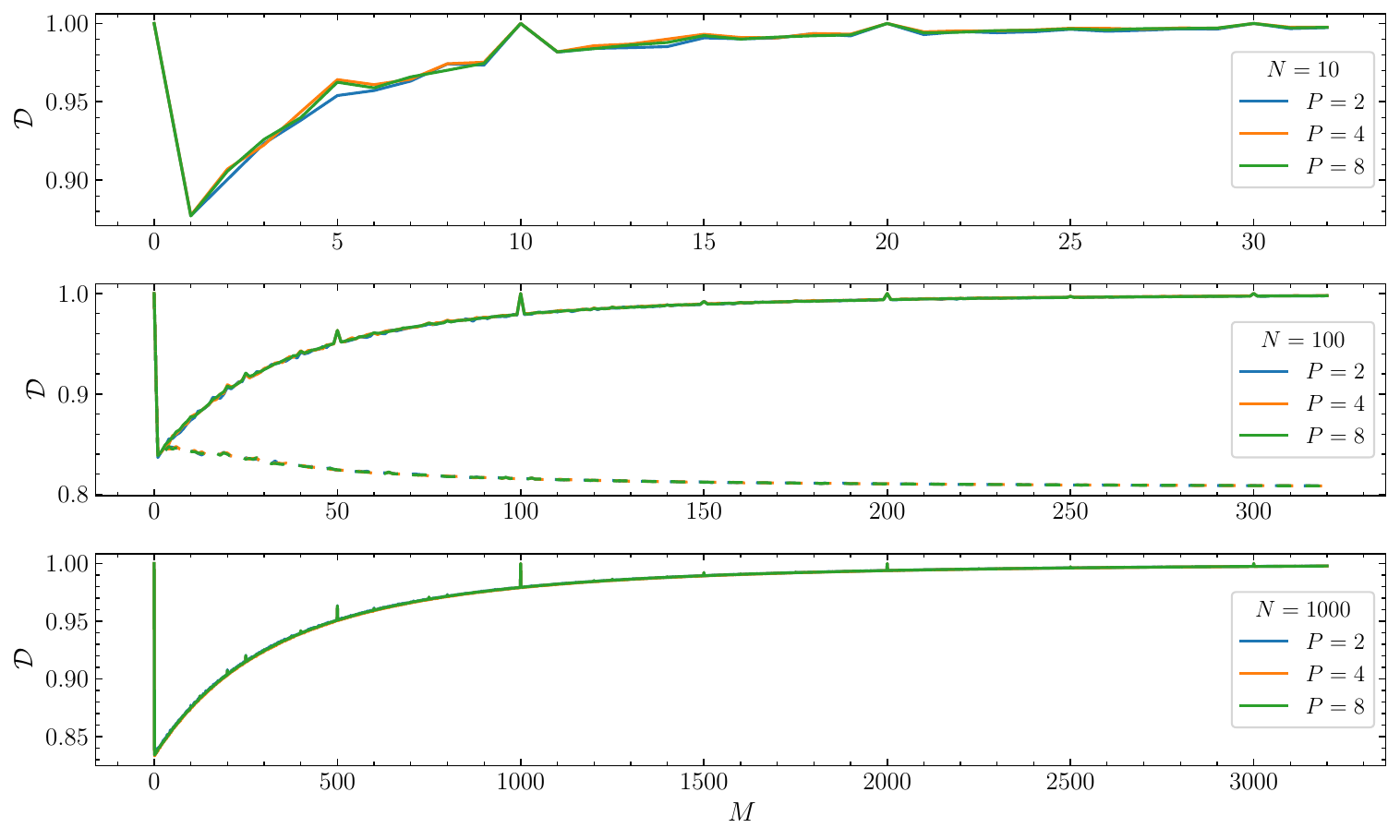}}
	\caption{
	The degree $\D$ of our LHS expansion as a function of the size $M$. The top, middle, and bottom panel shows results for an initial LHS set with $N=10$, $N=100$, and $N=1000$ samples, respectively. We consider three hypercube dimensions: $P=2$ (blue), $P=4$ (orange), and $P=8$ (green). The middle panel also considers the case of repeated unitary expansions, adding one sample $M$ times until the required expansion size is reached (dashed curves). To avoid highlighting particularly poor or favorable configurations, results are averaged over 100 Latin  hypercube realizations.}
	\label{fig:degree}
\end{figure*}
We now present some general properties of our expansion algorithm with a particular focus on the degree metric introduced in Sec.~\ref{sec:two}. Figure \ref{fig:degree} shows the degree $\D$ as a function of the expansion size $M$, for initial LHS sets with different numbers of samples $N$ and parameter-space dimensions $P$. For stability, results are averaged over multiple realizations. 

Lower values of $M$ correspond to a higher likelihood of overlaps, leading to a lower degree and thus a less space-filling eLHS. This effect diminishes as  $M$ increases; when 
$M>N$, the expansion dominates over the initial set, and the degree approaches 1. Figure \ref{fig:degree}  shows ``spikes'' with $\D=1$ taking place at multiples of the initial set size, i,e. $M=kN$ with $k\in\mathbb{N}$. This is not surprising: in those expansions, the existing boundaries of each bin remain unchanged and each initial interval is divided into $k$ subintervals. This implies that overlaps never occur, resulting in a perfect expansion.
Hints of this behavior were already presented in Refs.~\cite{SHEIKHOLESLAMI2017109, SALLABERRY20081047}, even though the authors only considered the case with $M=N$. %
 Less trivially, Fig.~\ref{fig:degree} shows that expansions with sizes  $M=\left(k+\frac{1}{2}\right)N$ exhibit higher degrees than adjacent values, although they do not reach $\D=1$.

We find that the LHS degree is largely independent of the number of dimensions $P$ and presents a self-similar behavior with $N$. In other words, $\D$ depends on the ratio $M/N$ and not on any of these two parameters independently. In particular, the degree is well predicted by the following phenomenological expression\footnote{More precisely, a least-square fit to the curves of Fig.~\ref{fig:degree} with ansatz $\D = 1 +a (b+M/N)^c$ returns $a=-0.167$, $b=1.01$, and $c=-2.99$.  }
\begin{equation}
\D = 1-\frac{1}{6\, (1+M/N)^{3}}\,,
\end{equation}
which, however, does not capture the high-$\D$ spikes described above.

Finally, the middle panel of Fig.~\ref{fig:degree} compares our results against repeated unitary expansions, that is, expanding the LHS $M$ times with one new sample at a time, instead of performing a direct one-step expansion of size $M$ as considered here. As expected, a unitary expansion strategy is highly suboptimal, as it leads to an overall increase in the likelihood of overlaps. Furthermore, it drastically increases the computational time by a factor $\propto M$. %

\subsection{Computational performance}

\begin{figure*}
	\centering
	\begin{subfigure}{.5\textwidth}
		\centering
		\includegraphics[width=0.9\columnwidth]{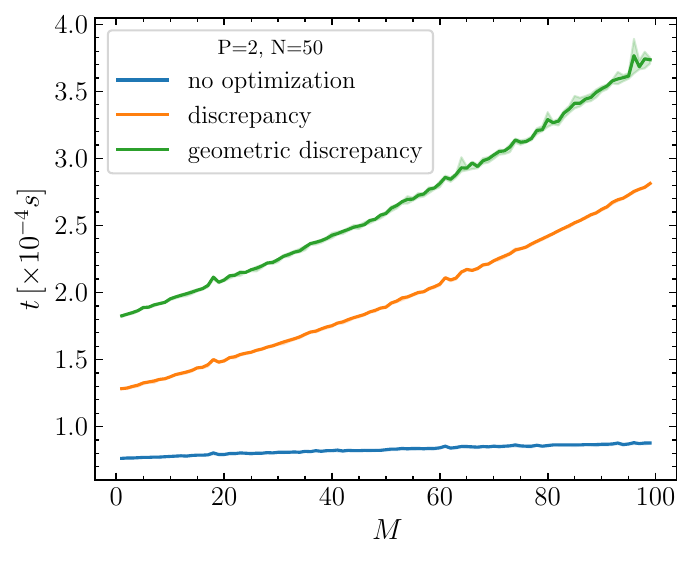}
		\label{subfig:1}
	\end{subfigure}%
	\begin{subfigure}{.5\textwidth}
		\centering
		\includegraphics[width=0.9\columnwidth]{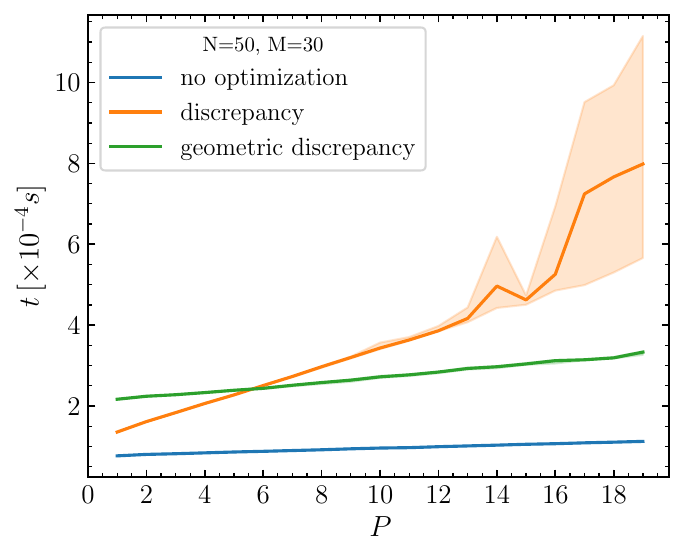}
		\label{subfig:2}
	\end{subfigure}
	\caption{Computational performance of the ``LHS in LHS’’ expansion algorithm. The left panel shows computational times for different expansion sizes $M$ of an initial LHS set with $N=50$ samples in $P=2$ dimensions. The right panel shows computational times for $N=50$ and $M=30$ when varying the number of hypercube dimensions $P$. Different curves represent variations of the expansion algorithm: no optimization in blue, discrepancy minimization in orange, and geometric discrepancy maximization in green. Solid curves show median values over 1000 LHS realizations, while the shaded regions encompass 90\% of the cases.
The time estimates reported in this figure were obtained by running \elhs in a single thread on an AMD EPYC Rome processor.
	}
	\label{fig:times}
\end{figure*}

Figure \ref{fig:times} illustrates the computational cost of \elhs as a function of $M$ and $P$. Just-in-time compilation via \numba significantly speeds up the computation, particularly when the expansion is performed without optimization. Unsurprisingly, the execution time increases when optimization is required. In particular, optimization degrades the algorithm’s scaling with the dimensionality of the hypercube.

Both of our optimization schemes show a similar trend as the number of samples grows, with discrepancy performing slightly better than geometric discrepancy. This scaling behavior can be explained as follows. The discrepancy focuses on uniformly filling the hypercube and quantifies the difference between the input distribution of points and the expected uniform coverage. Its evaluation requires comparing the fraction of samples in a hypercube subset with the fraction of volume it occupies, which depends heavily on $P$ and mildly on $M$. On the other hand, the geometric discrepancy estimates the minimum distance between any pair of points and is thus mainly affected by the number of samples $N+M$.

\section{Impact}

\label{sec:four}
LHS plays a major role in simulation design, as evidenced by the vast literature on the topic (for a review see Ref.~{\cite{2022arXiv220306334D}). The additional flexibility provided by our expansion algorithms broadens the use cases of this powerful tool. One interesting application is in training machine learning models. A model initially trained on a set of $N$ samples may fail to achieve the required accuracy. A common solution in machine learning is to increase the training set size or extend the coverage of the input space. An expansion algorithm for LHS can assist in selecting an appropriate distribution for new training samples while preserving existing information. A study on this topic is presented in Ref.~\cite{IORDANIS2025256}. To this end, expanded LHS might complement the usage of learning curves in machine learning \cite{mohr2024learning}. Quoting the specific field of research of some of us, gravitational-wave astronomy, two activities to which we plan to apply \elhs are the development of numerical-relativity surrogate models \cite{2019PhRvR...1c3015V, 2023PhRvD.108h4015B} and the interpolation of stellar-physics simulations \cite{2018PhRvD..98h3017T, 2022PhRvD.106j3013M}.

Our \elhs algorithm shares some similarities with the method presented in Ref.~\cite{Zhou2019}. The key differences are that our approach assumes fewer constraints on the final samples and directly targets the required number of additional samples $M$ while allowing for expansions that are not perfect Latin Hypercube sets.
Reference \cite{SHEIKHOLESLAMI2017109} also presented a related algorithm, which they dubbed Progressive LHS, supported by a public \python implementation. Their approach samples an LHS while also ensuring additional properties derived from Sliced Latin Hypercube sampling \cite{Qian01032012}. 
In particular, they construct a sequence of LHS slices whose progressive union still forms a Latin hypercube. The complete set obtained by the union of all slices constitutes a Latin hypercube that maximizes space-filling properties. Crucially, this remains a \textit{one-stage} sampling procedure, meaning that the size of the final Latin hypercube and the number of slices must be specified \textit{a priori}. Moreover, the algorithm permits sampling only those configurations in which the total number of samples is an integer multiple of the number of slices.
Our implementation is different; the regridding and ``LHS in LHS’’ operations we propose are arguably closer to the original procedure for distributing points via LHS and, by relaxing the requirement of a perfect Latin hypercube, we allow for expansions of any size.

\section{Conclusions}

In this paper, we presented a new algorithm for expanding an existing LHS while optimally preserving its space-filling properties. Our procedure leverages the flexibility of LHS by embedding a new LHS within the initial one. Loss of optimality arises from the potential occurrence of overlapping samples in the regridded space, as quantified by the LHS degree $\D$. 

Our new algorithm has broad applications in simulation design and machine learning, assisting in scenarios where a carefully planned initial simulation must later be expanded to meet targeted requirements.
However, the expansion strategy introduced in Sec.~\ref{sec:two} is specialized in sampling the uniform hypercube. Other LHS-like algorithms have been proposed to sample constrained spaces, handle non-uniform boundaries, and incorporate weighted samples (e.g. \cite{2009arXiv0909.0329P, 2006CG.....32.1378M, 2024arXiv240716567S}). Generalizing our implementation to these non-standard LHS techniques is left to future work.

Our implementation, \elhs, is available as an open-source package for the Python programming language, utilizing fast array manipulation and just-in-time compilation.

\section*{Acknowledgements}
\label{}

M.B., D.G., A.C. and M.F. are supported by
ERC Starting Grant No.~945155--GWmining, 
Cariplo Foundation Grant No.~2021-0555, 
MUR PRIN Grant No.~2022-Z9X4XS, 
MUR Grant ``Progetto Dipartimenti di Eccellenza 2023-2027'' (BiCoQ),
and the ICSC National Research Centre funded by NextGenerationEU. 
D.G. is supported by MSCA Fellowship
No.~101064542--StochRewind, MSCA Fellowship No.~101149270--ProtoBH, and MUR Young Researchers Grant No. SOE2024-0000125.
Computational work was performed at CINECA with allocations 
through INFN and Bicocca.

\bibliographystyle{elsarticle-num} 
\bibliography{lhssoftx_bibliography}

\end{document}